# Frequency conversion in topological plasmonic THz photo-mixer

Hamid Javadi

Submillimeter Wave Advanced Technology Group
Jet Propulsion Laboratory, Pasadena, CA 91109

We propose an alternative explanation for the observed coherent down-conversion in a laser-enabled plasmonic photo-mixer involving infrared/millimeter-wave/microwave photons. It has been demonstrated experimentally that a plasmonic gold grating deposited on low-temperature grown GaAs under illumination by a femto-second mode-locked pulsed laser, and simultaneously by a millimeter-wave source, outputs a microwave signal which is frequency-coherent with the comb laser's repetition frequency and millimeter-wave signal. We will offer a path toward full experimental characterization of the device

Here we will discuss a niche mixer design architecture for NASA applications that enables a wide range of space missions such as heterodyne receivers for astronomical observations, and high frequency resolution ultra-broad-band spectrometers for planetary observations. Wave-guide based heterodyne mixers have been the focus of development for many decades starting with whisker-contacted semiconductor Schottky diodes [1] transitioning to planar Schottky diodes [2] covering up to 2.4 THz [3]. Recently low-noise MMIC amplifiers (LNA) are routinely being employed as the receiver's front-end [4]. Cryogenic SIS junctions are employed for high sensitivity receivers [5]. Superconductor-to-normal-transition-based devices, and hot-electron bolometers offer other technological approaches (direct detection) for NASA applications [6]. Kinetic inductance superconductor-based detectors offer frequency spectrum multiplexing capability [7]. A laser-enabled plasmonic THz photo-mixer device is a new technology [8] that offers an alternative path to achieve high sensitivity / high frequency resolution receivers for NASA applications. Here, we will propose a new understanding of the physics of laser-enabled THz photo-mixer down-converter and offer directions for its further characterization and improvement. We extend the concept of topological bulk-edge correspondence from one-dimensional (1D) Su-Schrieffer-Heeger (SSH) tight-binding model [9] to device structure under consideration here. The design of the THz photo-mixer device is described in reference [10]: it is based on a one-dimensional (1D) plasmonic gold nanowire grating deposited on a Low-Temperature grown Gallium-Arsenide (LT-GaAs) substrate and placed in the apex of a THz antenna with contact leads for low-frequency signal pick-up (figure 1):

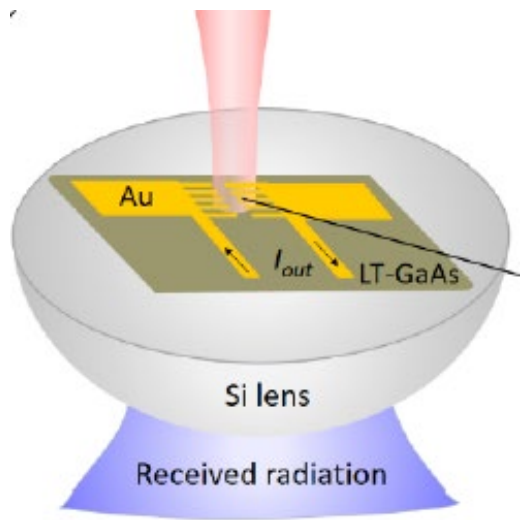

Fig. 1)- THz photo-mixer design [10]

when illuminated by a femto-second mode-locked pulsed laser (with 100 mW average power) [11] and a Gunn oscillator, it generates an intermediate frequency (IF) signal whose frequency is the difference between the mode-locked laser's comb pulse train @810 nm carrier frequency (regularly spaced in frequency by 75 MHz) and the Gunn oscillator's 96 GHz signal. The frequency-coherent down-converted 1 GHz IF frequency was detected and used for high-resolution frequency measurement of multiple millimeter-wave sources [10]. Here we offer a new theoretical understanding of the above-mentioned laser-enabled photo-mixer based on quantum-physics by invoking the concept of topological bulk-edge correspondence.

In their seminal paper, Su-Schrieffer-Heeger [9] presented a tight-binding model that describes the electronic transport dynamics for polyacetylene polymer chain where a $\pi$-bond between two neighboring Carbon atoms (double lines in the sketch below) supports two distinct, yet equivalent trans- configurations:

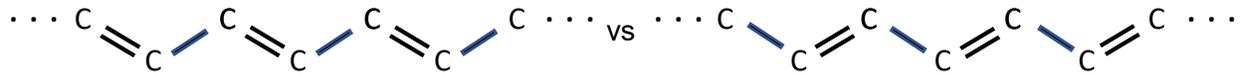

In its pristine, undoped state, polyacetylene is an insulator. Its doping by charged atoms adds electron or holes to the $\pi$- electron system (polaron) that can move along the polymer chain thus increasing its electrical conductivity. A highly doped polyacetylene can achieve very high conductivity approaching that of copper. Concepts such as 1D localization, hopping between 1D chains leading to delocalization and 3D conductivity have been investigated. Exotic species such as self-localized nonlinear excitations (topological solitons, polarons and bi-polarons) are further studied [12]. In polyacetylene, solitons can survive up to ambient temperatures but are prone to scattering by phonons, and trapping by localized impurities. Since theoretical inception of solitons in polyacetylene, SSH model has become the stepping stone for a new field of research studying the role of topology in quantum condensed matter physics.

Following Miranda et al [13], we extend the SSH model to plasmonic nanowire grating on LT-GaAs. The sketch in figure 2 represents the structure of a device similar to the one used in Wang et. al. [10] where the grating consists of gold nanowires (50 nm thick, 100 nm wide with 200 nm periodicity) deposited with Ti/N buffer layer (to establish ohmic contact for the contact leads) on LT-GaAs substrate. In region B of the device (figure 2) under light illumination, an incident photon with energy preferably above the band gap can generate an electron-hole pair that changes the conductivity of LT-GaAs substrate ($\sigma_B > \sigma_A$).

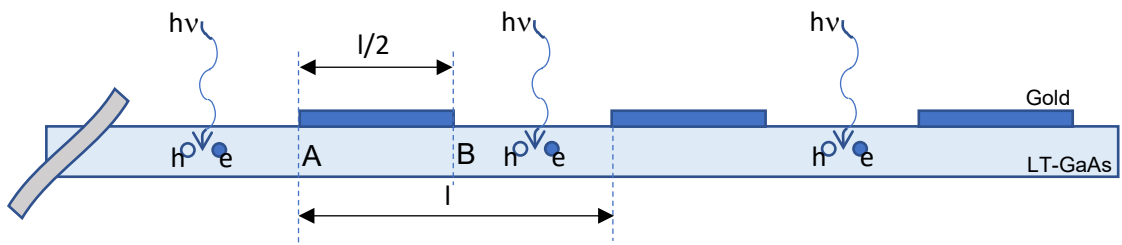

Fig. 2)- A 1D plasmonic nanowire grating structure showing charge particles generated under laser illumination

Similar to equation 19 of reference [13], we write total effective electronic Hamiltonian in momentum space (summing all wavenumbers $k$):

$$H = \sum_k [\psi_0^\dagger \; \xi_0^\dagger] \begin{bmatrix} 0 & J_A + J_B e^{-ikl} \\ J_A + J_B e^{ikl} & 0 \end{bmatrix} \begin{bmatrix} \psi_0 \\ \xi_0 \end{bmatrix}$$

where $J_A$, and $J_B$ are hopping amplitudes in the tight-binding model of sites A and B with site energies at zero volts. $\psi_0$, and $\xi_0$ are wavefunctions at sites A, and B reference points respectively. $k$ is Bloch wavenumber parameter, and $l$ is the period of 1D lattice's unit cell.

$$J_A = \frac{\sigma_A k_A}{\sin(k_A l/2)},\ J_B = \frac{\sigma_B k_B}{\sin(k_B l/2)},\ C = J_A \cos(k_A l/2) + J_B \cos(k_B l/2)$$

$k_A$ and $k_B$ are wavenumbers in the regions A and B, respectively. $C$ is the eigenvalue of the SSH model. Notice that in our model the two regions A and B have the same physical size ($l_A = l_B = l/2$) but different conductivity $\sigma_B > \sigma_A$.

Alternatively, our SSH model of plasmonic nanowire grating on LT-GaAs can be recast as composite right / left handed (CRLH) transmission line [14] whose incremental section ($\Delta x$) consists of inductors and capacitors in series / parallel (see figure 3):

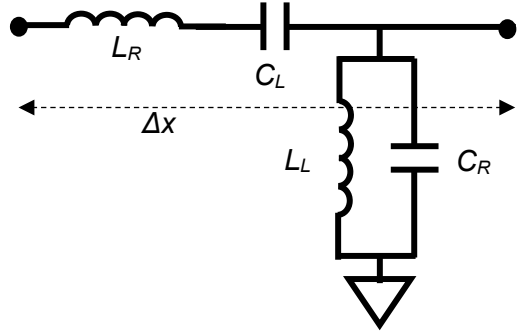

Fig. 3)- Composite right / left handed transmission line (CRLH) model of plasmonic nanowire grating supporting topological surface plasmon polariton [after 14]

The energy spectrum follows the relationship (similar to equation A16 of reference [14]):

$$[2 \sin(\frac{k \Delta x}{2})]^2 = \left(\frac{\omega}{\omega_c}\right)^2 + \frac{1}{\zeta \eta}\left(\frac{\omega_c}{\omega}\right)^2 - \frac{1}{\zeta} - \frac{1}{\eta}$$

where $\omega_c = \frac{1}{\sqrt{L_R C_R}}$ is the characteristic resonance frequency, $\zeta = \frac{C_L}{C_R}$, and $\eta = \frac{L_L}{L_R}$. The CRLH model considered here presents a gapped energy spectrum when the topological condition prevails [14] e.g.; at $k = 0$, a gap appears between $\omega_c/\sqrt{\eta}$ and $\omega_c/\sqrt{\zeta}$. The energy spectrum is very similar to the dispersion relation for a 1D linear array of metallic nanowires embedded in a dielectric media [15, 16]. The Yee-scheme [17] that is used for the construction of the finite-difference time-domain (FDTD) numerical analysis model has strong synergy with CRLH modelling. Traditionally one uses commercial numerical analysis tools (such as FDTD based Ansys Lumerical [18], or finite element model based Comsol [19]) to calculate the electromagnetic field profile within a 1D periodic array of nanoribbon grating on a substrate under illumination by a laser light [20] (figure 4).

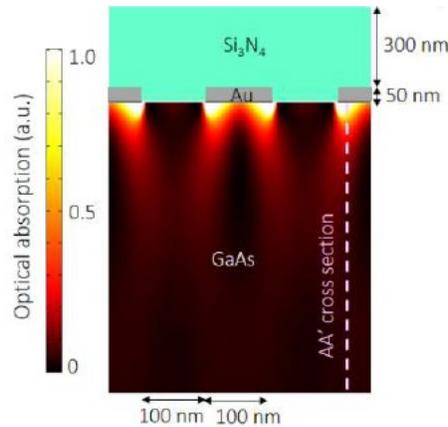

Fig. 4)- Numerical analysis (Comsol) of a 1D periodic array of gold grating embedded between two media (one unit cell is shown) showing electric field enhancement near the gold nanoribbon. LT-GaAs substrate is shown at the bottom and Si$_3$N$_4$ on top. A laser (wavelength of 784 nm) shines on the device. Bloch boundary conditions are applied to the side walls along horizontal axis [20].

Enhancement of the electric field near the metal nanostructure (figure 4) establishes the strong-coupling link between the laser-induced exciton in LT-GaAs substrate and the hybridized surface plasmon polariton wave riding on the 1D chain of nanowire grating.

An intense laser pulse shining on a LT-GaAs substrate generates excitons with an inhomogeneously broadened transition frequency as observed in linear absorption spectrum measurement at 10K [21] (in contrast to a sharp bound exciton resonance peak at 1.515 eV (~366 THz) for high-temperature grown GaAs).

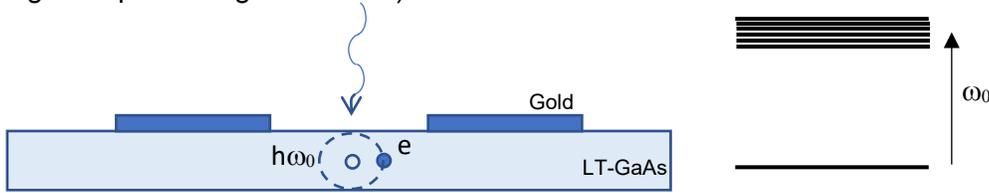

Fig. 6)- A graphical representation of an exciton in LT-GaAs substrate generated under intense Ti-Sapphire laser pulse, and the associated inhomogeneously broadened two-level system, with a distribution of closely-spaced excitation energies around ω₀. Ti-Sapphire laser reaches hundreds of KW pulse peak power and average power reaches 50 mW

Following Chen et. al. [22], going back to the quantum treatment of the grating, we can add an AC time-dependent hopping term of the form $V\cos(\Omega t)$ which will transform the tight binding Hamiltonian of a 1D lattice [with site energy $\omega_c$, time-independent hopping terms $h_k = J_A + J_B e^{-ik}$, and time-dependent hopping amplitude of $d_k = V(1 - e^{-ik})$] (similar to equation 6 of Chen et. al. [22]):

$$H_B(t) = \sum_k [\omega_c \widehat{a_k}^\dagger \widehat{a_k} + \omega_c \widehat{b_k}^\dagger \widehat{b_k} + (h_k \widehat{a_k}^\dagger \widehat{b_k} + \text{H.C.})] + \sum_k [2d_k \cos(\Omega t) \widehat{a_k}^\dagger \widehat{b_k} + \text{H.C.}]$$

where $\widehat{a_k}/\widehat{a_k}^\dagger$, $\widehat{b_k}/\widehat{b_k}^\dagger$ are annihilation/creation operators for quantum species in sites A, and B respectively, and H.C. stands for hermetian conjugate. The time-dependent eigenfunctions and eigenvalues of this problem can be solved using Quantum Toolbox in Python (QuTiP): An open-source Python framework for the dynamics of open quantum systems [23]. For periodicity of $T = 2\pi/\Omega$, under Floquet formalism, the analysis yields quasi-energy spectrum composed of sidebands separated by a period of Ω. Following Chen et. al. [22], in figure 7 we depict this spectrum for a finite 1D lattice with Open Boundary Condition with the following choice of parameters: $J_A = 1$, $\Omega/J_A = 5$, $V/J_A = 0.2$. In figure 7, one sees quasi-energy spectrum of a finite chain with N = 30 unit cells as a function of ratio $J_B/J_A$. Starting at $J_B/J_A \cong 1$, a topological $0 - $ gap edge state is formed (seen at the lower and upper rails of the spectrum) which is then joined by a topological $\pi - $ gap edge state at $J_B/J_A \cong 1.5$ (seen as a straight line at the middle of butterfly figure shaped spectrum). The $\pi - $ gap edge state disappears at $J_B/J_A \cong 3.5$

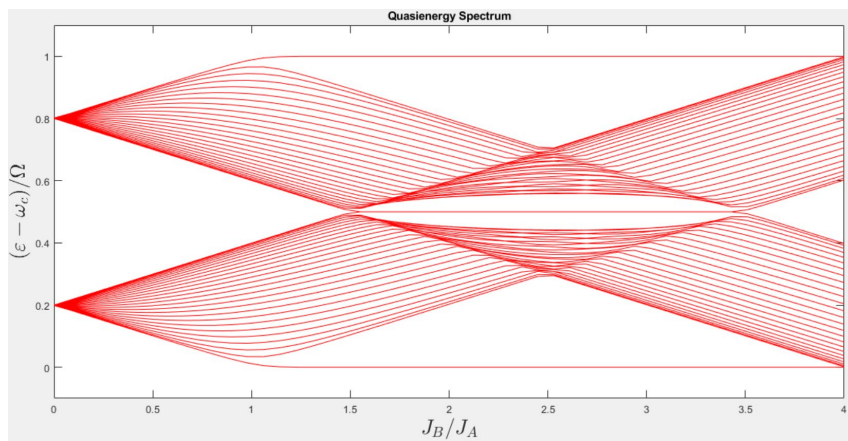

Fig. 7)- Quasi-energy spectrum for a finite 1D lattice (N=27) with Open Boundary Condition [22]

Dmytruk et al [24] considered a similar configuration as of the above; where a 1D SSH lattice was placed inside a single-mode photonic resonator. They studied the interplay between the topological phase of the matter with quantum light field. They showed that similar to the role of $J_B/J_A$ in transitioning from a trivial to a topological phase, at high coupling strength, the quantum field of the single-mode resonator can drive a trivial SSH chain to topological phase and preserve its chiral symmetry. Wave vector $k$ needs to be replaced by $k + (g/\sqrt{L}) * (\hat{a}_k + \hat{a}_k^\dagger)$ where g is coupling factor, $L$ is the length of SSH chain and $\hat{a}_k$ is the light quanta of the single-mode resonator [24]. In particular, the quasi-energy spectrum plot in figure 7 is exactly replicated for different values of $J_B/J_A$.

In figure 8, following Chen et. al. [22], we depict site occupancy; the occupation probability for sites A and B on the 1D lattice model as a function of $J_B/J_A$ taking values of 1.0, 2.0, 4.0 in panels from left to right (blue/red for site A/B):

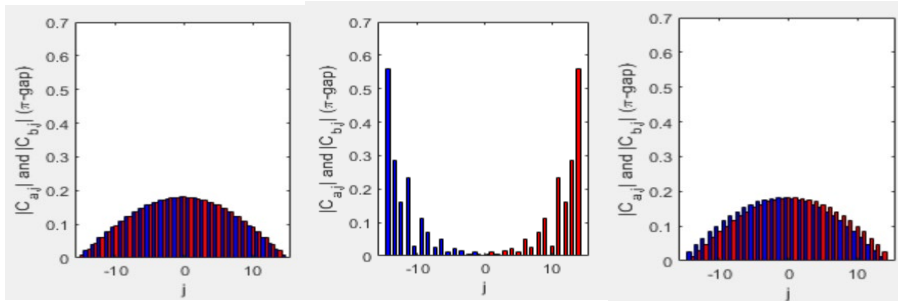

Fig. 8)- Progression of site-occupancy (vertical axis) as a function of site's location (horizontal axis - $\frac{L}{2}$ to $\frac{L}{2}$) for $J_B/J_A$ = 1.0, 2.0, 4.0 across horizontal panels (from left to right) showing formation of topological $\pi -$ gap edge state in the middle panel for $J_B/J_A$ = 2.0 (site A: blue, site B: red) [22].

in the middle panel, one sees how the $\pi -$ gap edge state is concentrated on the edges of the lattice chain at t = 0. As time passes, site occupancy switches between A/B on the left/right of the lattice (see below and also figure 9) [22].

The $\pi -$ gap edge state appearing for middle panel in figure 8 ($J_B/J_A = 2.0$) has time-dependency of period of $T = 2\pi/\Omega$ (where $\Omega$ is the frequency of the drive) switching from site A / B concentration on the right / left hand side of the 1D lattice $(t_0 = 0)/(t_0 = T/2)$; it can be seen in figure 9 (following Chen et. al. [22]):

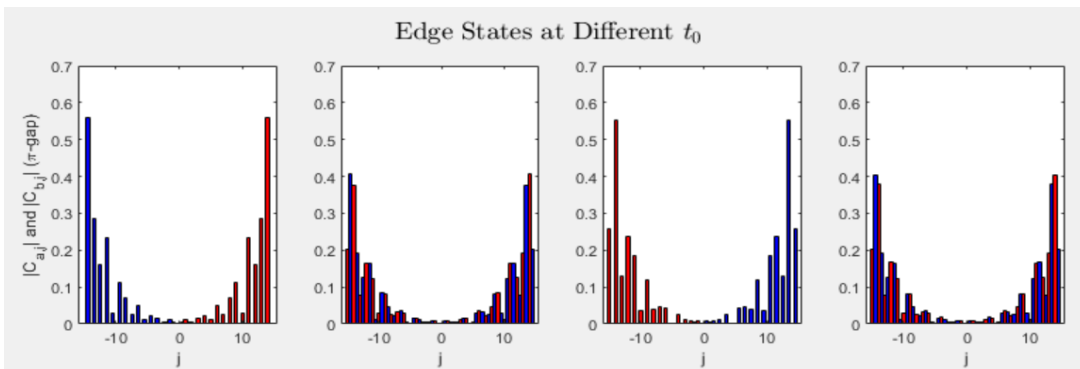

Fig. 9)- Progression of topological $\pi -$ gap edge state ($J_B/J_A$= 2) for $t_0$ = 0 / $\frac{T}{4}$ / $\frac{T}{2}$ / $\frac{3T}{4}$ left-to-right (site A: blue, site B: red) [22].

panels at t = 0 and t = T/2 show that $\pi -$ gap edge state site occupancy switches from site A → B with time. When the device is illuminated by an intense femto-second laser beam, two distinct excitons could form within the LT-GaAs substrate with frequencies $\omega_1 = \omega_c + \Delta_1$ and $\omega_2 = \omega_c + \Delta_2$ (figure 10). Following Chen et. al. [22], in figure 11 we show lattice site occupancy or two excitons generated at sites 1B (left panels) and 29B (right panels) of N = 30 sites lattice chain as $J_B/J_A$ increases from 1.0 to 2.0 (top row to bottom row of panels):

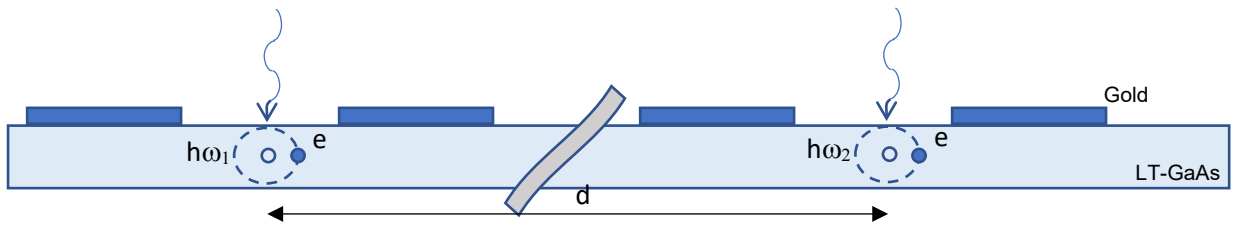

Fig. 10)- A graphical representation of two distinct excitons (separated by distance d) generated in LT-GaAs substrate under intense Ti-Sapphire comb laser pulses.

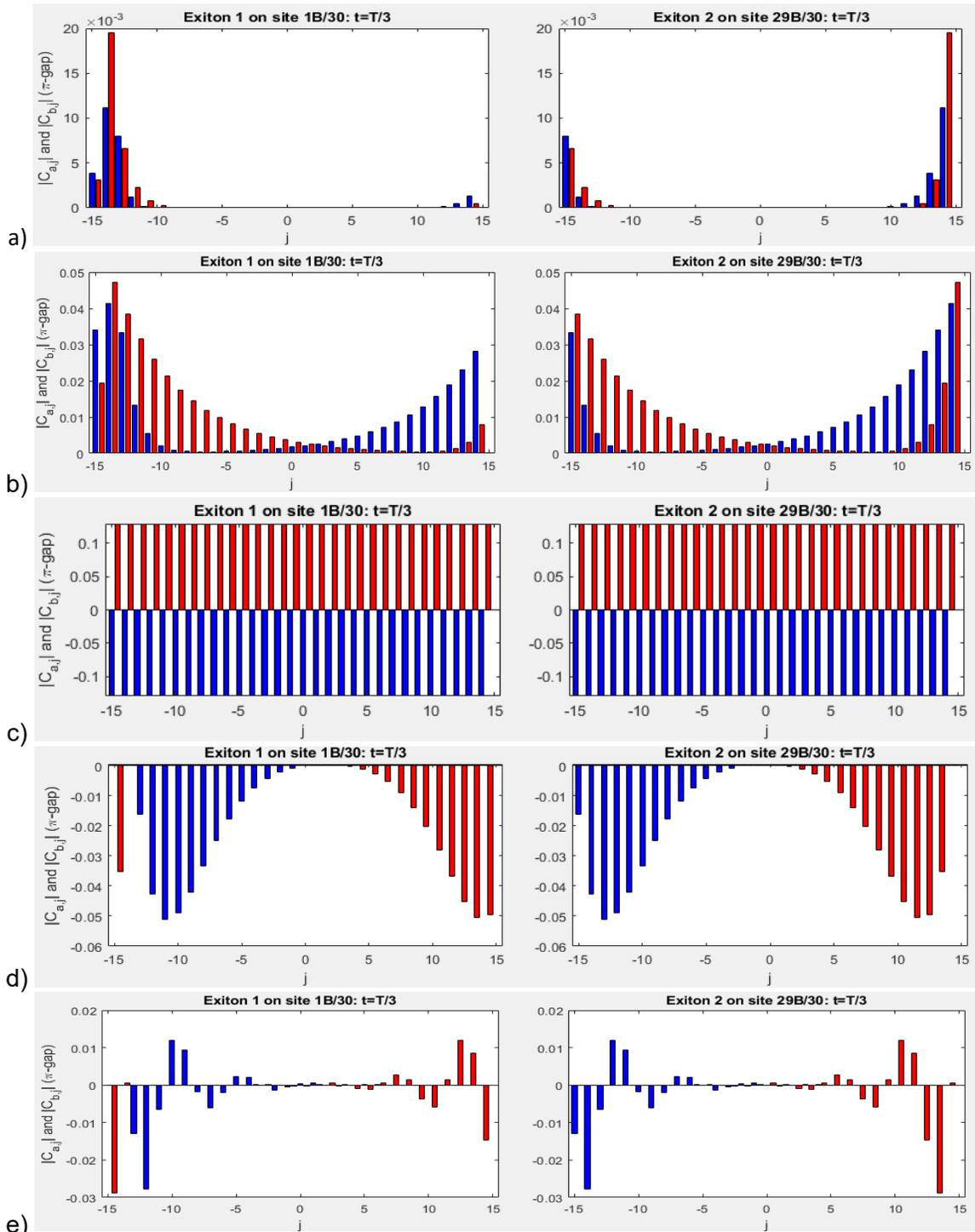

Fig. 11)- Site occupancies when excitons 1 and 2 are coupled to sites 1B and 29B, respectively. Excitons are bound to $\pi-$gap edge states for $J_B/J_A = 1.0, 1.45, 1.5, 1.55, 2.0$ (panels from top) [22].

In general, the site occupancy for these two cases (excitons at 1B and 29B) is similar (left and right panels in figure 11) thus the overlap of the two wavefunctions is optimum. it is clear that at the start of $\pi-$ gap edge state formation, when $J_B/J_A = 1.5$, site occupancy is not solely concentrated on two edges of the lattice chain but rather it is spread fully across the lattice. The calculated dipole-dipole interaction increases as one approaches $J_B/J_A = 1.5$ where it diverges.

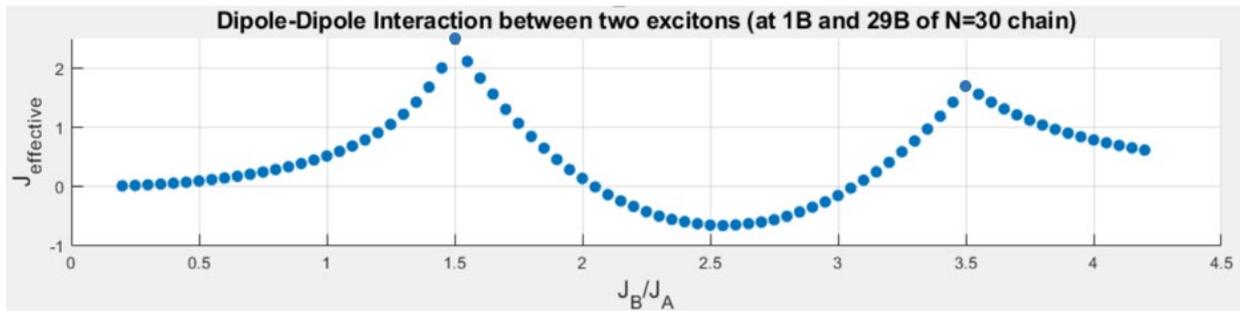

Fig. 12)- Variation of evaluated dipole-dipole interaction for two excitons located at opposite ends of the SSH chain indicates critical scaling at the appearance/disappearance of the topological $\pi-$ gap edge state at as $J_B/J_A = 1.5/3.5$

Surface Plasmon Polariton (SPP) is the optical excitation at the interface of plasmonic nanostructure and the dielectric media above (air) and the dielectric media below (substrate). SPP possess an inherently weak nonlinearity and short propagation distance (few microns). An exciton with large dipole moment could interact strongly with SPP forming hybrid exciton-SPP modes when their energy level splitting is more than each individual line-widths. The hybrid exciton/SPP offers enhanced nonlinearity that can be utilized in applications such as the one considered here. Here, the excitons become bound to topological $\pi-$ gap edge states of plasmonic grating and interact with each other via quasi-particles [25] riding on the 1D plasmonic grating.

So far, we have drawn a parallel between topological SSH model and 1D plasmonic grating on LT-GaAs substrate. We have extended the concept of two quantum emitters interacting via virtual photons on a dielectric waveguide to interaction between two excitons embedded in a 1D SPP nanostructure. One further questions whether the interaction between such excitons can survive disorder and thermal fluctuations (at room temperature) similar to the topological edge states.

One is then obliged to provide experimental evidences supporting such claims. Our purpose here is to encourage researchers in the field to consider the possibility of such parallelism (as discussed in this paper) and to pursue further investigation into substantiating the expressed hypothesis. At the end of this paper, we show preliminary steps taken by us to do the same.

Following Chen et. al. [22], we look at time-dependence of the exchange transfer rate for $\pi-$ gap edge state for $J_B/J_A = 1.475, 1.49, 1.5, 1.51, 1.525$ and its frequency components. As can be seen in figure 13 middle row for $J_B/J_A = 1.5$, the strong-peaks frequency differences are:

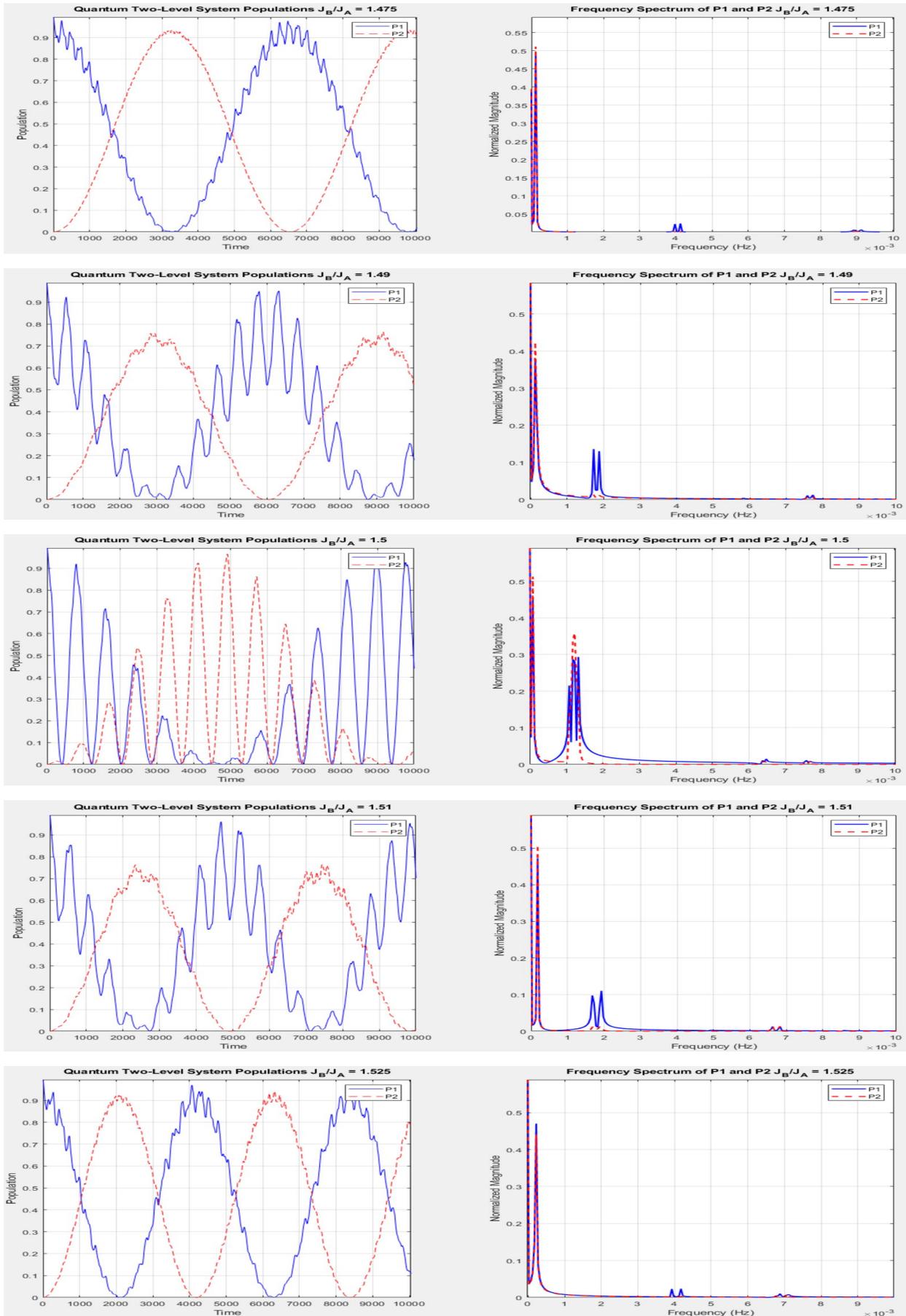

Fig. 13)- Exchange rate as a function of time (left panel) and its associated Fourier-transform spectra (right panel) for $J_B/J_A = 1.475, 1.49, 1.5, 1.51, 1.525$ arranged from top to bottom [22].

as it can be seen in the figure 13, two (2) satellite twin-peaks move toward zero frequency as $J_B/J_A$ get close to 1.5. At $J_B/J_A = 1.5,$ there are three frequency peaks.

The plasmonic photo-mixer in reference [10] has a large parameter space spanning frequency ranges from IF (~1 GHz), to millimeter-wave (mm-wave ~100 GHz), to optics (810 nm $\cong$ 370 THz). To properly evaluate any device, one needs to measure the characteristics S-parameter for all ports including their return loss and port-to-port insertion loss. At first approximation, the 6x6 S-parameter matrix representing the network configuration of the device is constructed of three (3) central 2x2 matrices, one for each distinct IF/mm-wave/optics frequency band (input/output). We have designed a mechanical jig (figures 14, 15) to couple in the 100 GHz mm-wave signal via waveguide-to-microstripline transition based on [26]. We have optimized the circuit's design, and obtained acceptable return-losses and insertion losses for the 1-GHz and 100-GHz signals (figure 16). The device is placed at the apex of a bow-tie antenna where the 100 GHz signal is concentrated. Millimeter-wave signal is provided via WR-10 waveguides (colored brown in figure 14). The detected IF signal is routed out to sma connectors (colored blue in figure 14) via low-pass filters to avoid leakage of the higher frequencies.

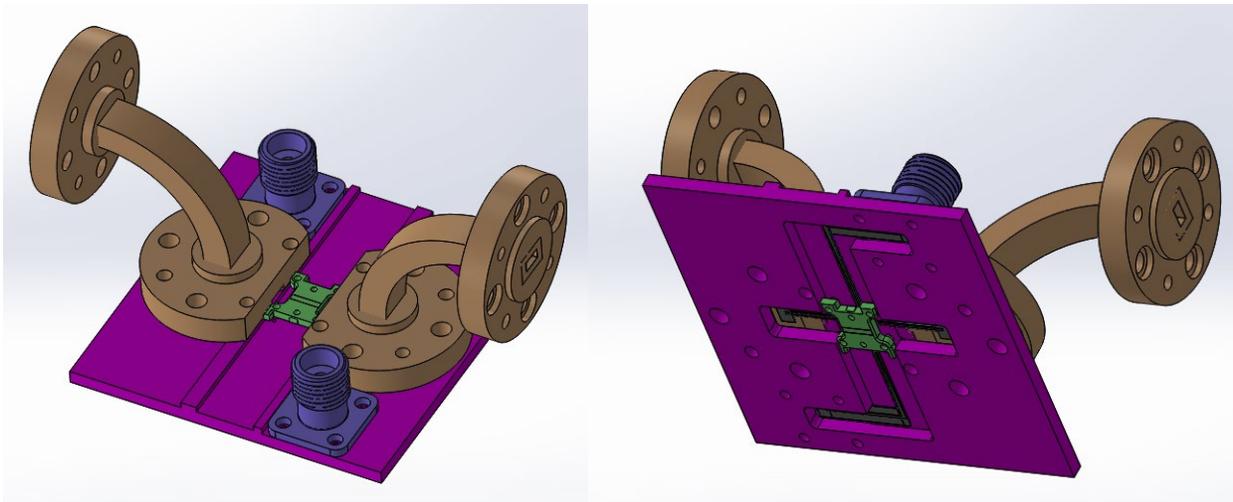

Figure 14- Mechanical drawing of the test jig to characterize the proposed laser-based THz photo-mixer. There are two WR-10 waveguides that couple 100 GHz in- and out- of the device via waveguide-to-microstripline transition. Two sma connectors are for IF frequency (~1 GHz) coupled to the device via low pass filters. The laser light is also brought in via a fiber laser and focused to the device via a lensed optical connector. A second fiber connector picks up the transmitted light.

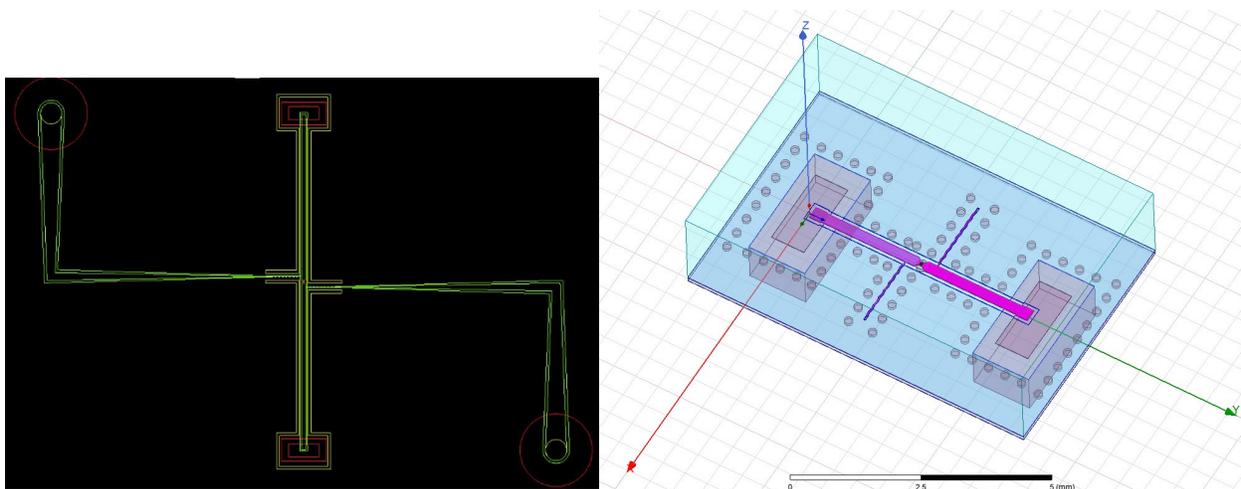

Figure 15- The circuit on the left is made of gold microstripline deposited on thin alumina substarte. On the right we see the transmission line carrying 100 GHz signal from WR-10 input waveguide to WR-10 output waveguide.

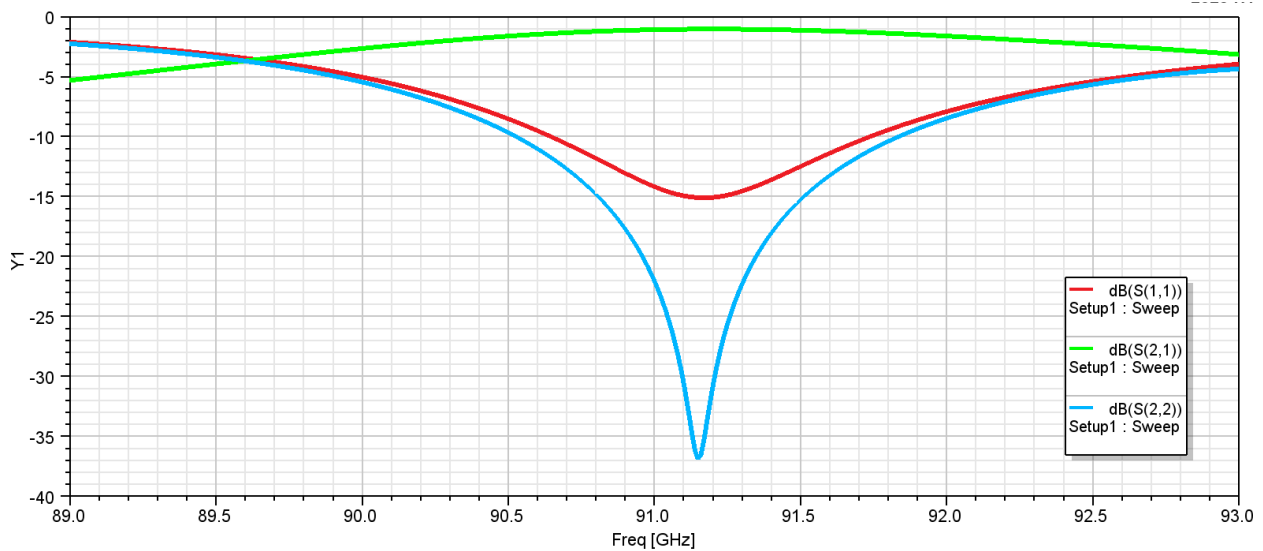

Figure 16- Jig's WR-10 waveguide→micro-stripline→waveguide return/insertion loss

Two (2) femtosecond pulsed lasers are combined in a single optical fiber and transferred to the device via a specialized parallel optical fibers bundle [27] (green colored piece in figure 14) with a total internal reflection mirror and a linear array of dielectric lenses delivering the laser beam perpendicular to the planar device and focused on the linear array of point targets (figure 17).

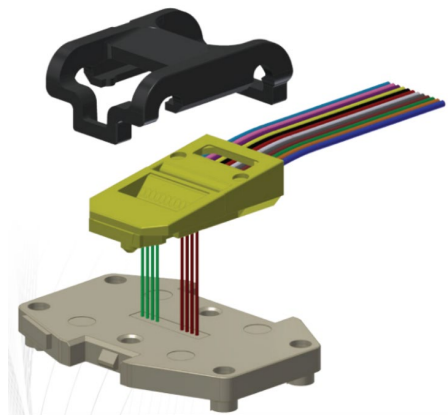

Figure 17- artist drawing of USCONEC PRIZM [27] optical delivery system LightTurn connector with typical insertion loss 1.2 dB @850 nm.

We have not analyzed the performance of the optical system but made sure that one can measure the scalar transmission and reflection power of the laser bouncing off the device. The jig is machined and is ready for integration of the substrate, device, and laser delivery system. The circuit has a large footprint. To achieve a micro-stripline built to the spec to high tolerance, we need a repeatable and accurate manufacturing technique. That forces us to limit the size of the alumina substrate. The circuit thus has to be built in multiple sections and connected via ribbon welds. Each end of the micro-stripline has a pickup probe that enters the waveguide opening. An alternative technique to construct the mm-wave circuit is via laser etching (Protolaser 200; resolution scan field: $2\ \mu m$, repeatability: $\pm 2\ \mu m$) [28] of metal cladding on commercially available RT/duroid [29]. Unfortunately, this technique is in its infancy and has a low yield, yet some good results are obtained through trial and error [30]. Further advancement involves replacing bulky Ti-Sapphire pulsed lasers with the first chip-scale integrated miniaturized version [31] sold by [32].